%Paper: astro-ph/9305024
%From: viotti%astrom.hepnet@Lbl.Gov
%Date: Wed, 19 May 93 09:17:51 PDT

%%%%%%%%%%%%%%%%%%%%%%%%%%%%%%%%%%%%%%%%%%%%%%%%%%%%%%%%%%%%%%%%%%%%%%%
% This is AA.CMM, the plain TeX macro package
% (CM version) from Springer-Verlag
% for the Astronomy and Astrophysics Main Journal
% Version 2.0 as of 25 Feb 1991
%
% Test for recursive or multiple loading of Springer macro packages
\def\SpringerMacroPackageNameATest{AA}%
\let\next\relax
\ifx\SpringerMacroPackageNameA\undefined
  \message{Loading the \SpringerMacroPackageNameATest\space
           macro package from Springer-Verlag...}%
\else
  \ifx\SpringerMacroPackageNameA\SpringerMacroPackageNameATest
    \message{\SpringerMacroPackageNameA\space macro package
             from Springer-Verlag already loaded.}%
    \let\next\endinput
  \else
    \message{DANGER: \SpringerMacroPackageNameA\space from
             Springer-Verlag already loaded, will try to proceed.}%
  \fi
\fi
\next
\def\SpringerMacroPackageNameA{AA}%
% now call all the sub-macros
% indention of equations
\newskip\mathindent      \mathindent=0pt
% \titlea
\newskip\tabefore \tabefore=20dd plus 10pt minus 5pt      % space above
\newskip\taafter  \taafter=10dd                           % space below
% \titleb
\newskip\tbbeforeback    \tbbeforeback=-20dd              % corrective space to
%%a \titlea
\newskip\tbbefore        \tbbefore=17pt plus 7pt minus3pt % spaceabove
\newskip\tbafter         \tbafter=8pt                     % space below
% \titlec
\newskip\tcbeforeback    \tcbeforeback=-3pt               % corrective space to
%%a \titleb
\advance\tcbeforeback by -10dd                            % corrective space to
%%a \titleb
\newskip\tcbefore        \tcbefore=10dd plus 5pt minus 1pt% space above
\newskip\tcafter         \tcafter=6pt                     % space below
% \titled
\newskip\tdbeforeback    \tdbeforeback=-3pt                  % corrective space
%%to a \titlec
\advance\tdbeforeback by -10dd                               % corrective space
%%to a \titlec
\newskip\tdbefore        \tdbefore=10dd plus 4pt minus 1pt   % space above
% \petit
\newskip\petitsurround
\petitsurround=6pt\relax
% \ack
\newskip\ackbefore      \ackbefore=10dd plus 5pt             % space above
\newskip\ackafter       \ackafter=6pt                        % space below
% indention of lists
\newdimen\itemindent    \newdimen\itemitemindent
\itemindent=1.5em       \itemitemindent=2\itemindent
 \font \tatt            = cmbx10 scaled \magstep3
 \font \tats            = cmbx10 scaled \magstep1
 \font \tamt            = cmmib10 scaled \magstep3
 \font \tams            = cmmib10 scaled \magstep1
 \font \tamss           = cmmib10
 \font \tast            = cmsy10 scaled \magstep3
 \font \tass            = cmsy10 scaled \magstep1
 \font \tbtt            = cmbx10 scaled \magstep2
 \font \tbmt            = cmmib10 scaled \magstep2
 \font \tbst            = cmsy10 scaled \magstep2
\catcode`@=11    % use @ as a normal character
\vsize=23.5truecm
\hoffset=-1true cm
\voffset=-1true cm
\normallineskip=1dd
\normallineskiplimit=0dd
\newskip\ttglue%
\def\ifundefin@d#1#2{%
\expandafter\ifx\csname#1#2\endcsname\relax}
\def\getf@nt#1#2#3#4{%
\ifundefin@d{#1}{#2}%
\global\expandafter\font\csname#1#2\endcsname=#3#4%
\fi\relax
}
\newfam\sffam
\newfam\scfam
\def\makesize#1#2#3#4#5#6#7{%
 \getf@nt{rm}{#1}{cmr}{#2}%
 \getf@nt{rm}{#3}{cmr}{#4}%
 \getf@nt{rm}{#5}{cmr}{#6}%
 \getf@nt{mi}{#1}{cmmi}{#2}%
 \getf@nt{mi}{#3}{cmmi}{#4}%
 \getf@nt{mi}{#5}{cmmi}{#6}%
 \getf@nt{sy}{#1}{cmsy}{#2}%
 \getf@nt{sy}{#3}{cmsy}{#4}%
 \getf@nt{sy}{#5}{cmsy}{#6}%
 \skewchar\csname mi#1\endcsname ='177
 \skewchar\csname mi#3\endcsname ='177
 \skewchar\csname mi#5\endcsname ='177
 \skewchar\csname sy#1\endcsname ='60
 \skewchar\csname sy#3\endcsname='60
 \skewchar\csname sy#5\endcsname='60
\expandafter\def\csname#1size\endcsname{%
 \normalbaselineskip=#7
 \normalbaselines
 \setbox\strutbox=\hbox{\vrule height0.75\normalbaselineskip%
    depth0.25\normalbaselineskip width0pt}%
 \textfont0=\csname rm#1\endcsname
 \scriptfont0=\csname rm#3\endcsname
 \scriptscriptfont0=\csname rm#5\endcsname
    \def\oldstyle{\fam1\csname mi#1\endcsname}%
 \textfont1=\csname mi#1\endcsname
 \scriptfont1=\csname mi#3\endcsname
 \scriptscriptfont1=\csname mi#5\endcsname
 \textfont2=\csname sy#1\endcsname
 \scriptfont2=\csname sy#3\endcsname
 \scriptscriptfont2=\csname sy#5\endcsname
 \textfont3=\tenex\scriptfont3=\tenex\scriptscriptfont3=\tenex
   \def\rm{%
 \fam0\csname rm#1\endcsname%
   }%
   \def\it{%
 \getf@nt{it}{#1}{cmti}{#2}%
 \textfont\itfam=\csname it#1\endcsname
 \fam\itfam\csname it#1\endcsname
   }%
   \def\sl{%
 \getf@nt{sl}{#1}{cmsl}{#2}%
 \textfont\slfam=\csname sl#1\endcsname
 \fam\slfam\csname sl#1\endcsname}%
   \def\bf{%
 \getf@nt{bf}{#1}{cmbx}{#2}%
 \getf@nt{bf}{#3}{cmbx}{#4}%
 \getf@nt{bf}{#5}{cmbx}{#6}%
 \textfont\bffam=\csname bf#1\endcsname
 \scriptfont\bffam=\csname bf#3\endcsname
 \scriptscriptfont\bffam=\csname bf#5\endcsname
 \fam\bffam\csname bf#1\endcsname}%
   \def\tt{%
 \getf@nt{tt}{#1}{cmtt}{#2}%
 \textfont\ttfam=\csname tt#1\endcsname
 \fam\ttfam\csname tt#1\endcsname
 \ttglue=.5em plus.25em minus.15em
   }%
  \def\sf{%
\getf@nt{sf}{#1}{cmss}{10 at #2pt}%
\textfont\sffam=\csname sf#1\endcsname
\fam\sffam\csname sf#1\endcsname}%
   \def\sc{%
 \getf@nt{sc}{#1}{cmcsc}{10 at #2pt}%
 \textfont\scfam=\csname sc#1\endcsname
 \fam\scfam\csname sc#1\endcsname}%
\rm }}
\makesize{IXf}{9}{VIf}{6}{Vf}{5}{10.00dd}
\def\normalsize{\IXfsize
\def\sf{%
   \getf@nt{sf}{IXf}{cmss}{9}%
   \getf@nt{sf}{VIf}{cmss}{10 at 6pt}%
   \getf@nt{sf}{Vf}{cmss}{10 at 5pt}%
   \textfont\sffam=\csname sfIXf\endcsname
   \scriptfont\sffam=\csname sfVIf\endcsname
   \scriptscriptfont\sffam=\csname sfVf\endcsname
   \fam\sffam\csname sfIXf\endcsname}%
}%
\newfam\mibfam
\def\mib{%
   \getf@nt{mib}{IXf}{cmmib}{10 at9pt}%
   \getf@nt{mib}{VIf}{cmmib}{10 at6pt}%
   \getf@nt{mib}{Vf}{cmmib}{10 at5pt}%
   \textfont\mibfam=\csname mibIXf\endcsname
   \scriptfont\mibfam=\csname mibVIf\endcsname
   \scriptscriptfont\mibfam=\csname mibVf\endcsname
   \fam\mibfam\csname mibIXf\endcsname}%
\makesize{Xf}{10}{VIf}{6}{Vf}{5}{10.00dd}
\Xfsize
\it\bf\tt\rm

\def\tentt{\ttXf}

\normalsize
\it\bf\tt\sf\mib\rm
\def\boldmath{\textfont1=\mibIXf \scriptfont1=\mibVIf
\scriptscriptfont1=\mibVf}
\newdimen\fullhsize
\newcount\verybad \verybad=1010
\let\lr=L%
\fullhsize=40cc
\hsize=19.5cc
\def\fullline{\hbox to\fullhsize}
\def\makefootline{\baselineskip=10dd \fullline{\the\footline}}
\def\makeheadline{\vbox to 0pt{\vskip-22.5pt
            \fullline{\vbox to 8.5pt{}\the\headline}\vss}\nointerlineskip}
\hfuzz=2pt
\vfuzz=2pt
\tolerance=1000
\abovedisplayskip=3 mm plus6pt minus 4pt
\belowdisplayskip=3 mm plus6pt minus 4pt
\abovedisplayshortskip=0mm plus6pt
\belowdisplayshortskip=2 mm plus4pt minus 4pt
\parindent=1.5em
\newdimen\stdparindent\stdparindent\parindent
\frenchspacing
\nopagenumbers
\predisplaypenalty=600        % Make a page break before a display harder
\displaywidowpenalty=2000     % and even harder for a widow display.
\def\widowsandclubs#1{\global\verybad=#1
   \global\widowpenalty=\the\verybad1      % default: 10101
   \global\clubpenalty=\the\verybad2  }    % default: 10102
\widowsandclubs{1010}
\def\paglay{\headline={{\normalsize\hsize=.75\fullhsize\ifnum\pageno=1
\vbox{\hrule\line{\vrule\kern3pt\vbox{\kern3pt
\hbox{\bf A\&A manuscript no.}
\hbox{(will be inserted by hand later)}
\kern3pt\hrule\kern3pt
\hbox{\bf Your thesaurus codes are:}
\hbox{\rightskip=0pt plus3em\advance\hsize by-7pt
\vbox{\bf\noindent\ignorespaces\the\THESAURUS}}
\kern3pt}\hfil\kern3pt\vrule}\hrule}
\rlap{\quad\AALogo}\hfil
\else\normalsize\ifodd\pageno\hfil\folio\else\folio\hfil\fi\fi}}}
\makesize{VIIIf}{8}{VIf}{6}{Vf}{5}{9.00dd}
      \getf@nt{sf}{VIIIf}{cmss}{8}%
      \getf@nt{sf}{VIf}{cmss}{10 at 6pt}%
      \getf@nt{sf}{Vf}{cmss}{10 at 5pt}%
      \getf@nt{mib}{VIIIf}{cmmib}{10 at 8pt}%
      \getf@nt{mib}{VIf}{cmmib}{10 at 6pt}%
      \getf@nt{mib}{Vf}{cmmib}{10 at 5pt}%
\VIIIfsize\it\bf\tt\rm
\normalsize
\def\petit{\VIIIfsize
   \def\sf{%
      \getf@nt{sf}{VIIIf}{cmss}{8}%
      \getf@nt{sf}{VIf}{cmss}{10 at 6pt}%
      \getf@nt{sf}{Vf}{cmss}{10 at 5pt}%
      \textfont\sffam=\csname sfVIIIf\endcsname
      \scriptfont\sffam=\csname sfVIf\endcsname
      \scriptscriptfont\sffam=\csname sfVf\endcsname
      \fam\sffam\csname sfVIIIf\endcsname
}%
\def\mib{%
   \getf@nt{mib}{VIIIf}{cmmib}{10 at 8pt}%
   \getf@nt{mib}{VIf}{cmmib}{10 at 6pt}%
   \getf@nt{mib}{Vf}{cmmib}{10 at 5pt}%
   \textfont\mibfam=\csname mibVIIIf\endcsname
   \scriptfont\mibfam=\csname mibVIf\endcsname
   \scriptscriptfont\mibfam=\csname mibVf\endcsname
   \fam\mibfam\csname mibIXf\endcsname}%
\def\boldmath{\textfont1=\mibVIIIf\scriptfont1=\mibVIf
\scriptscriptfont1=\mibVf}%
\let\bfIXf=\bfVIIIf
 \if Y\REFEREE \normalbaselineskip=2\normalbaselineskip
 \normallineskip=2\normallineskip\fi
 \setbox\strutbox=\hbox{\vrule height7pt depth2pt width0pt}%
 \normalbaselines\rm}%
\def\begpet{\vskip\petitsurround
\bgroup\petit}%  Beginn eines Paragraphen in petit
\def\endpet{\vskip\petitsurround
\egroup}%  Ende eines Paragraphen in petit

 \let  \tatss           = \bfXf
 \let  \tasss           = \syXf
 \let  \tbts            = \bfXf
 \let  \tbtss           = \bfVIIIf
 \let  \tbms            = \tamss
 \let  \tbmss           = \mibVIIIf
 \let  \tbss            = \syXf
 \let  \tbsss           = \syVIIIf
\def\newline{\hfill\break}% makes a new line in the text :)
\def\rahmen#1{\vbox{\hrule\line{\vrule\vbox to#1true
cm{\vfil}\hfil\vrule}\vfil\hrule}}
\let\ts=\thinspace
\def\,{\relax\ifmmode\mskip\thinmuskip\else\thinspace\fi}
\def\unvskip{%
   \ifvmode
      \ifdim\lastskip=0pt
      \else
         \vskip-\lastskip
      \fi
   \fi}
\newtoks\eq\newtoks\eqn
\newdimen\mathhsize
\def\calcmathhsize{\mathhsize=\hsize
\advance\mathhsize by-\mathindent}
\calcmathhsize
\def\eqalign#1{\null\vcenter{\openup\jot\m@th
  \ialign{\strut\hfil$\displaystyle{##}$&$\displaystyle{{}##}$\hfil
      \crcr#1\crcr}}}
\def\displaylines#1{{}$\displ@y
\hbox{\vbox{\halign{$\@lign\hfil\displaystyle##\hfil$\crcr
    #1\crcr}}}${}}
\def\eqalignno#1{{}$\displ@y
  \hbox{\vbox{\halign
to\mathhsize{\hfil$\@lign\displaystyle{##}$\tabskip\z@skip
    &$\@lign\displaystyle{{}##}$\hfil\tabskip\centering
    &\llap{$\@lign##$}\tabskip\z@skip\crcr
    #1\crcr}}}${}}
\def\leqalignno#1{{}$\displ@y
\hbox{\vbox{\halign
to\mathhsize{\qquad\hfil$\@lign\displaystyle{##}$\tabskip\z@skip
    &$\@lign\displaystyle{{}##}$\hfil\tabskip\centering
    &\kern-\mathhsize\rlap{$\@lign##$}\tabskip\hsize\crcr
    #1\crcr}}}${}}
\def\generaldisplay{%
\ifeqno
       \ifleqno\leftline{$\displaystyle\the\eqn\quad\the\eq$}%
       \else\noindent\kern\mathindent\hbox to\mathhsize{$\displaystyle
             \the\eq\hfill\the\eqn$}%
       \fi
\else
       \kern\mathindent
       \hbox to\mathhsize{$\displaystyle\the\eq$\hss}%
\fi
\global\eq={}\global\eqn={}}%
\newif\ifeqno\newif\ifleqno
\everydisplay{\displaysetup}
\def\displaysetup#1$${\displaytest#1\eqno\eqno\displaytest}
% look for equation numbers
\def\displaytest#1\eqno#2\eqno#3\displaytest{%
\if!#3!\ldisplaytest#1\leqno\leqno\ldisplaytest
\else\eqnotrue\leqnofalse\eqn={#2}\eq={#1}\fi
\generaldisplay$$}
\def\ldisplaytest#1\leqno#2\leqno#3\ldisplaytest{\eq={#1}%
\if!#3!\eqnofalse\else\eqnotrue\leqnotrue\eqn={#2}\fi}
\newcount\eqnum\eqnum=0% register
\def\autnum{\global\advance\eqnum by 1\relax{\rm(\the\eqnum)}}
\newdimen\lindent
\lindent=\stdparindent

\def\litemitem{\par\noindent\hbox to\lindent{\hfil}%
               \hangindent=2\lindent\ltextindent}
\def\ltextindent#1{\hbox to\lindent{#1\hss}\ignorespaces}
\def\set@item@mark#1{\llap{#1\enspace}\ignorespaces}
\ifx\undefined\mathhsize
   \def\item{\par\noindent
   \hangindent\itemindent\hangafter=0
   \set@item@mark}
   \def\itemitem{\par\noindent\advance\mathhsize by-\itemitemindent
   \hangindent\itemitemindent\hangafter=0
   \set@item@mark}
\else
   \def\item{\par\noindent\advance\mathhsize by-\itemindent
   \hangindent\itemindent\hangafter=0
   \everypar={\global\mathhsize=\hsize
   \global\advance\mathhsize by-\mathindent
   \global\everypar={}}\set@item@mark}
   \def\itemitem{\par\noindent\advance\mathhsize by-\itemitemindent
   \hangindent\itemitemindent\hangafter=0
   \everypar={\global\mathhsize=\hsize
   \global\advance\mathhsize by-\mathindent
   \global\everypar={}}\set@item@mark}
\fi
\newcount\the@end \global\the@end=0
\newbox\springer@macro \setbox\springer@macro=\vbox{}
\def\typeset{\setbox\springer@macro=\vbox{\begpet\noindent
   This article was processed by the author using
   Sprin\-ger-Ver\-lag \TeX{} A\&A macro package 1991.\par
   \egroup}\global\the@end=1}
\outer\def\bye{\bigskip\typeset
\sterne=1\ifx\speciali\undefined
\else
  \loop\smallskip\noindent special character No\number\sterne:
    \csname special\romannumeral\sterne\endcsname
    \advance\sterne by 1\relax
    \ifnum\sterne<11\relax
  \repeat
\fi
\if R\lr\null\fi\vfill\supereject\end}
\def\AALogo{\setbox254=\hbox{ ASTROPHYSICS }%
\vbox{\baselineskip=10dd\hrule\hbox{\vrule\vbox{\kern3pt
\hbox to\wd254{\hfil ASTRONOMY\hfil}
\hbox to\wd254{\hfil AND\hfil}\copy254
\hbox to\wd254{\hfil\number\day.\number\month.\number\year\hfil}
\kern3pt}\vrule}\hrule}}
\def\figure#1#2{\medskip\noindent{\petit{\bf Fig.\ts#1.\
}\ignorespaces#2\par}}
\expandafter \newcount \csname c@Tl\endcsname
    \csname c@Tl\endcsname=0
\expandafter \newcount \csname c@Tm\endcsname
    \csname c@Tm\endcsname=0
\expandafter \newcount \csname c@Tn\endcsname
    \csname c@Tn\endcsname=0
\expandafter \newcount \csname c@To\endcsname
    \csname c@To\endcsname=0
\expandafter \newcount \csname c@Tp\endcsname
    \csname c@Tp\endcsname=0
\expandafter \newcount \csname c@fn\endcsname
    \csname c@fn\endcsname=0
\def \stepc#1    {\global
    \expandafter
    \advance
    \csname c@#1\endcsname by 1}
\def \resetcount#1    {\global
    \csname c@#1\endcsname=0}
\def\@nameuse#1{\csname #1\endcsname}
\def\arabic#1{\@arabic{\@nameuse{c@#1}}}
\def\@arabic#1{\ifnum #1>0 \number #1\fi}
 \def \aTa  { \goodbreak
     \bgroup
     \par
 \textfont0=\tatt \scriptfont0=\tats \scriptscriptfont0=\tatss
 \textfont1=\tamt \scriptfont1=\tams \scriptscriptfont1=\tamss
 \textfont2=\tast \scriptfont2=\tass \scriptscriptfont2=\tasss
     \baselineskip=17dd\lineskiplimit=0pt\lineskip=0pt
     \rightskip=0pt plus4cm
     \pretolerance=10000
     \noindent
     \tatt}
 \def \eTa{\vskip10pt\egroup
     \noindent
     \ignorespaces}
 \def \aTb{\goodbreak
     \bgroup
     \par
 \textfont0=\tbtt \scriptfont0=\tbts \scriptscriptfont0=\tbtss
 \textfont1=\tbmt \scriptfont1=\tbms \scriptscriptfont1=\tbmss
 \textfont2=\tbst \scriptfont2=\tbss \scriptscriptfont2=\tbsss
     \baselineskip=13dd\lineskip=0pt\lineskiplimit=0pt
     \rightskip=0pt plus4cm
     \pretolerance=10000
     \noindent
     \tbtt}
 \def \eTb{\vskip10pt
     \egroup
     \noindent
     \ignorespaces}
\newcount\section@penalty  \section@penalty=0
\newcount\subsection@penalty  \subsection@penalty=0
\newcount\subsubsection@penalty  \subsubsection@penalty=0
\def\titlea#1{\par\stepc{Tl}
    \resetcount{Tm}
    \bgroup
       \normalsize
       \bf \rightskip 0pt plus4em
       \pretolerance=20000
       \boldmath
       \setbox0=\vbox{\vskip\tabefore
          \noindent
          \arabic{Tl}.\
          \ignorespaces#1
          \vskip\taafter}
       \dimen0=\ht0\advance\dimen0 by\dp0
       \advance\dimen0 by 2\baselineskip
       \advance\dimen0 by\pagetotal
       \ifdim\dimen0>\pagegoal
          \ifdim\pagetotal>\pagegoal
          \else\eject\fi\fi
       \vskip\tabefore
       \penalty\section@penalty \global\section@penalty=-200
       \global\subsection@penalty=10007
       \noindent
       \arabic{Tl}.\
       \ignorespaces#1
       \vskip\taafter
    \egroup
    \nobreak
    \parindent=0pt
    \let\lasttitle=A%
\everypar={\parindent=\stdparindent
    \penalty\z@\let\lasttitle=N\everypar={}}%
    \ignorespaces}
\def\titleb#1{\par\stepc{Tm}
    \resetcount{Tn}
    \if N\lasttitle\else\vskip\tbbeforeback\fi
    \bgroup
       \normalsize
       \raggedright
       \pretolerance=10000
       \it
       \setbox0=\vbox{\vskip\tbbefore
          \normalsize
          \raggedright
          \pretolerance=10000
          \noindent \it \arabic{Tl}.\arabic{Tm}.\ \ignorespaces#1
          \vskip\tbafter}
       \dimen0=\ht0\advance\dimen0 by\dp0\advance\dimen0 by 2\baselineskip
       \advance\dimen0 by\pagetotal
       \ifdim\dimen0>\pagegoal
          \ifdim\pagetotal>\pagegoal
          \else \if N\lasttitle\eject\fi \fi\fi
       \vskip\tbbefore
       \if N\lasttitle \penalty\subsection@penalty \fi
       \global\subsection@penalty=-100
       \global\subsubsection@penalty=10007
       \noindent \arabic{Tl}.\arabic{Tm}.\ \ignorespaces#1
       \vskip\tbafter
    \egroup
    \nobreak
    \let\lasttitle=B%
    \parindent=0pt
    \everypar={\parindent=\stdparindent
       \penalty\z@\let\lasttitle=N\everypar={}}%
       \ignorespaces}
\def\titlec#1{\par\stepc{Tn}
    \resetcount{To}
    \if N\lasttitle\else\vskip\tcbeforeback\fi
    \bgroup
       \normalsize
       \raggedright
       \pretolerance=10000
       \setbox0=\vbox{\vskip\tcbefore
          \noindent
          \arabic{Tl}.\arabic{Tm}.\arabic{Tn}.\
          \ignorespaces#1\vskip\tcafter}
       \dimen0=\ht0\advance\dimen0 by\dp0\advance\dimen0 by 2\baselineskip
       \advance\dimen0 by\pagetotal
       \ifdim\dimen0>\pagegoal
           \ifdim\pagetotal>\pagegoal
           \else \if N\lasttitle\eject\fi \fi\fi
       \vskip\tcbefore
       \if N\lasttitle \penalty\subsubsection@penalty \fi
       \global\subsubsection@penalty=-50
       \noindent
       \arabic{Tl}.\arabic{Tm}.\arabic{Tn}.\
       \ignorespaces#1\vskip\tcafter
    \egroup
    \nobreak
    \let\lasttitle=C%
    \parindent=0pt
    \everypar={\parindent=\stdparindent
       \penalty\z@\let\lasttitle=N\everypar={}}%
       \ignorespaces}
\def\titled#1{\par\stepc{To}
    \resetcount{Tp}
    \if N\lasttitle\else\vskip\tdbeforeback\fi
    \vskip\tdbefore
    \bgroup
       \normalsize
       \if N\lasttitle \penalty-50 \fi
       \it \noindent \ignorespaces#1\unskip\
    \egroup\ignorespaces}
\def\begref#1{\par
   \unvskip
   \goodbreak\vskip\tabefore
   {\noindent\bf\ignorespaces#1%
   \par\vskip\taafter}\nobreak\let\INS=N}
\def\ref{\if N\INS\let\INS=Y\else\goodbreak\fi
   \hangindent\stdparindent\hangafter=1\noindent\ignorespaces}
\def\endref{\goodbreak}% Ende der Referenzen
\def\acknow#1{\par
   \unvskip
   \vskip\tcbefore
   \noindent{\it Acknowledgements\/}. %
   \ignorespaces#1\par
   \vskip\tcafter}
\def\appendix#1{\vskip\tabefore
    \vbox{\noindent{\bf Appendix #1}\vskip\taafter}%
    \global\eqnum=0\relax
    \nobreak\noindent\ignorespaces}
\let\REFEREE=N
\newbox\refereebox
\setbox\refereebox=\vbox
to0pt{\vskip0.5cm\fullline{\hrulefill\tentt\lower0.5ex
\hbox{\kern5pt referee's copy\kern5pt}\hrulefill}\vss}%
\def\refereelayout{\let\REFEREE=M\footline={\copy\refereebox}
    \message{|A referee's copy will be produced}\par
    \if N\lr\else\if R\lr \onecolumn \fi \let\lr=N \topskip=10pt\fi}

\def\utw{\smash{\rlap{\lower5pt\hbox{$\sim$}}}}
\def\udtw{\smash{\rlap{\lower6pt\hbox{$\approx$}}}}

 %reelle Zahlen
 %natuerliche Zahlen

\def\bbbc{{\mathchoice {\setbox0=\hbox{$\displaystyle\rm C$}\hbox{\hbox
to0pt{\kern0.4\wd0\vrule height0.9\ht0\hss}\box0}}
{\setbox0=\hbox{$\textstyle\rm C$}\hbox{\hbox
to0pt{\kern0.4\wd0\vrule height0.9\ht0\hss}\box0}}
{\setbox0=\hbox{$\scriptstyle\rm C$}\hbox{\hbox
to0pt{\kern0.4\wd0\vrule height0.9\ht0\hss}\box0}}
{\setbox0=\hbox{$\scriptscriptstyle\rm C$}\hbox{\hbox
to0pt{\kern0.4\wd0\vrule height0.9\ht0\hss}\box0}}}}
\def\bbbq{{\mathchoice {\setbox0=\hbox{$\displaystyle\rm Q$}\hbox{\raise
0.15\ht0\hbox to0pt{\kern0.4\wd0\vrule height0.8\ht0\hss}\box0}}
{\setbox0=\hbox{$\textstyle\rm Q$}\hbox{\raise
0.15\ht0\hbox to0pt{\kern0.4\wd0\vrule height0.8\ht0\hss}\box0}}
{\setbox0=\hbox{$\scriptstyle\rm Q$}\hbox{\raise
0.15\ht0\hbox to0pt{\kern0.4\wd0\vrule height0.7\ht0\hss}\box0}}
{\setbox0=\hbox{$\scriptscriptstyle\rm Q$}\hbox{\raise
0.15\ht0\hbox to0pt{\kern0.4\wd0\vrule height0.7\ht0\hss}\box0}}}}
\def\bbbt{{\mathchoice {\setbox0=\hbox{$\displaystyle\rm
T$}\hbox{\hbox to0pt{\kern0.3\wd0\vrule height0.9\ht0\hss}\box0}}
{\setbox0=\hbox{$\textstyle\rm T$}\hbox{\hbox
to0pt{\kern0.3\wd0\vrule height0.9\ht0\hss}\box0}}
{\setbox0=\hbox{$\scriptstyle\rm T$}\hbox{\hbox
to0pt{\kern0.3\wd0\vrule height0.9\ht0\hss}\box0}}
{\setbox0=\hbox{$\scriptscriptstyle\rm T$}\hbox{\hbox
to0pt{\kern0.3\wd0\vrule height0.9\ht0\hss}\box0}}}}
\def\bbbs{{\mathchoice
{\setbox0=\hbox{$\displaystyle     \rm S$}\hbox{\raise0.5\ht0\hbox
to0pt{\kern0.35\wd0\vrule height0.45\ht0\hss}\hbox
to0pt{\kern0.55\wd0\vrule height0.5\ht0\hss}\box0}}
{\setbox0=\hbox{$\textstyle        \rm S$}\hbox{\raise0.5\ht0\hbox
to0pt{\kern0.35\wd0\vrule height0.45\ht0\hss}\hbox
to0pt{\kern0.55\wd0\vrule height0.5\ht0\hss}\box0}}
{\setbox0=\hbox{$\scriptstyle      \rm S$}\hbox{\raise0.5\ht0\hbox
to0pt{\kern0.35\wd0\vrule height0.45\ht0\hss}\raise0.05\ht0\hbox
to0pt{\kern0.5\wd0\vrule height0.45\ht0\hss}\box0}}
{\setbox0=\hbox{$\scriptscriptstyle\rm S$}\hbox{\raise0.5\ht0\hbox
to0pt{\kern0.4\wd0\vrule height0.45\ht0\hss}\raise0.05\ht0\hbox
to0pt{\kern0.55\wd0\vrule height0.45\ht0\hss}\box0}}}}
\def\bbbz{{\mathchoice {\hbox{$\sf\textstyle Z\kern-0.4em Z$}}
{\hbox{$\sf\textstyle Z\kern-0.4em Z$}}
{\hbox{$\sf\scriptstyle Z\kern-0.3em Z$}}
{\hbox{$\sf\scriptscriptstyle Z\kern-0.2em Z$}}}}
\def\diameter{{\ifmmode\oslash\else$\oslash$\fi}}

\def\vec#1{{\boldmath
\textfont0=\bfIXf\scriptfont0=\bfVIf\scriptscriptfont0=\bfVf
\ifmmode
\mathchoice{\hbox{$\displaystyle#1$}}{\hbox{$\textstyle#1$}}
{\hbox{$\scriptstyle#1$}}{\hbox{$\scriptscriptstyle#1$}}\else
$#1$\fi}}
\def\tens#1{\ifmmode
\mathchoice{\hbox{$\displaystyle\sf#1$}}{\hbox{$\textstyle\sf#1$}}
{\hbox{$\scriptstyle\sf#1$}}{\hbox{$\scriptscriptstyle\sf#1$}}\else
$\sf#1$\fi}
\newcount\sterne \sterne=0
\newdimen\fullhead
{\catcode`@=11    % use @ as a normal character
\def\newtoks{\alloc@5\toks\toksdef\@cclvi}
\outer\gdef\makenewtoks#1{\newtoks#1#1={ ????? }}}
\makenewtoks\DATE
\makenewtoks\MAINTITLE
\makenewtoks\SUBTITLE
\makenewtoks\AUTHOR
\makenewtoks\INSTITUTE
\makenewtoks\ABSTRACT
\makenewtoks\KEYWORDS
\makenewtoks\THESAURUS
\makenewtoks\OFFPRINTS
\newlinechar=`\| %
\let\INS=N%
{\catcode`\@=\active
\gdef@#1{\if N\INS $^{#1}$\else\if
E\INS\hangindent0.5\stdparindent\hangafter=1%
\noindent\hbox to0.5\stdparindent{$^{#1}$\hfil}\let\INS=Y\ignorespaces
\else\par\hangindent0.5\stdparindent\hangafter=1
\noindent\hbox to0.5\stdparindent{$^{#1}$\hfil}\ignorespaces\fi\fi}%
}%
\def\mehrsterne{\global\advance\sterne by1\relax}%
\def\footnoterule{\kern-3pt\hrule width 2true cm\kern2.6pt}% Trennlinie
\def\makeOFFPRINTS#1{\bgroup\normalsize
       \hsize=19.5cc
       \baselineskip=10dd\lineskiplimit=0pt\lineskip=0pt
       \def\textindent##1{\noindent{\it Send offprint
          requests to\/}: }\relax
       \vfootnote{nix}{\ignorespaces#1}\egroup}
\def\makesterne{\count254=0\loop\ifnum\count254<\sterne
\advance\count254 by1\star\repeat}
\def\FOOTNOTE#1{\bgroup
       \ifhmode\unskip\fi
       \mehrsterne$^{\makesterne}$\relax
       \normalsize
       \hsize=19.5cc
       \baselineskip=10dd\lineskiplimit=0pt\lineskip=0pt
       \def\textindent##1{\noindent\hbox
       to\stdparindent{##1\hss}}\relax
       \vfootnote{$^{\makesterne}$}{\ignorespaces#1}\egroup}
\def\fonote#1{\ifhmode\unskip\fi
       \mehrsterne$^{\the\sterne}$\bgroup
       \normalsize
       \hsize=19.5cc
       \def\textindent##1{\noindent\hbox
       to\stdparindent{##1\hss}}\relax
       \vfootnote{$^{\the\sterne}$}{\ignorespaces#1}\egroup}
\def\missmsg#1{\message{|Missing #1 }}
\def\tstmiss#1#2#3#4#5{%
\edef\test{\the #1}%
\ifx\test\missing%
  #2\relax%  message
  #3%   action if missing
\else
  \ifx\test\missingi%
    #2\relax%  message
    #3%   action if missing
  \else #4%  action if existing
  \fi
\fi
#5%   action at any rate
}%
\def\maketitle{\paglay%
\def\missing{ ????? }%
\def\missingi{ }%
{\parskip=0pt\relax
\setbox0=\vbox{\hsize=\fullhsize\null\vskip2truecm
\tstmiss%
  {\MAINTITLE}%
  {}%
  {\global\MAINTITLE={MAINTITLE should be given}}%
  {}%
  {%   write MAINTITLE:
   \aTa\ignorespaces\the\MAINTITLE\eTa}%
\tstmiss%
  {\SUBTITLE}%
  {}%
  {}%
  {%   write SUBTITLE:
   \aTb\ignorespaces\the\SUBTITLE\eTb}%
  {}%
\tstmiss%
  {\AUTHOR}%
  {}%
  {\AUTHOR={Name(s) and initial(s) of author(s) should be given}}
  {}%
  {%   write AUTHOR:
\noindent{\bf\ignorespaces\the\AUTHOR\vskip4pt}}%
\tstmiss%
  {\INSTITUTE}%
  {}%
  {\INSTITUTE={Address(es) of author(s) should be given.}}%
  {}%
  {%   write INSTITUTE:
   \let\INS=E
\noindent\ignorespaces\the\INSTITUTE\vskip10pt}%
\tstmiss%
  {\DATE}%
  {}%
  {\DATE={$[$the date of receipt and acceptance should be inserted
later$]$}}%
  {}%
  {%   write DATE:
{\noindent\ignorespaces\the\DATE\vskip21pt}\bf A}%
}%
\global\fullhead=\ht0\global\advance\fullhead by\dp0
\global\advance\fullhead by10pt\global\sterne=0
{\hsize=19.5cc\null\vskip2truecm
\tstmiss%
  {\OFFPRINTS}%
  {}%
  {}%
  {\makeOFFPRINTS{\the\OFFPRINTS}}%
  {}%
\hsize=\fullhsize
\tstmiss%
  {\MAINTITLE}%
  {\missmsg{MAINTITLE}}%
  {\global\MAINTITLE={MAINTITLE should be given}}%
  {}%
  {%   write MAINTITLE:
   \aTa\ignorespaces\the\MAINTITLE\eTa}%
\tstmiss%
  {\SUBTITLE}%
  {}%
  {}%
  {%   write SUBTITLE:
   \aTb\ignorespaces\the\SUBTITLE\eTb}%
  {}%
\tstmiss%
  {\AUTHOR}%
  {\missmsg{name(s) and initial(s) of author(s)}}%
  {\AUTHOR={Name(s) and initial(s) of author(s) should be given}}
  {}%
  {%   write AUTHOR:
\noindent{\bf\ignorespaces\the\AUTHOR\vskip4pt}}%
\tstmiss%
  {\INSTITUTE}%
  {\missmsg{address(es) of author(s)}}%
  {\INSTITUTE={Address(es) of author(s) should be given.}}%
  {}%
  {%   write INSTITUTE:
   \let\INS=E
\noindent\ignorespaces\the\INSTITUTE\vskip10pt}%
\catcode`\@=12
\tstmiss%
  {\DATE}%
  {\message{|The date of receipt and acceptance should be inserted
later.}}%
  {\DATE={$[$the date of receipt and acceptance should be inserted
later$]$}}%
  {}%
  {%   write DATE:
{\noindent\ignorespaces\the\DATE\vskip21pt}}%
}%
\tstmiss%
  {\THESAURUS}%
  {\message{|Thesaurus codes are not given.}}%
  {\global\THESAURUS={missing; you have not inserted them}}%
  {}%
  {}%
\if M\REFEREE\let\REFEREE=Y
\normalbaselineskip=2\normalbaselineskip
\normallineskip=2\normallineskip\normalbaselines\fi
\tstmiss%
  {\ABSTRACT}%
  {\missmsg{ABSTRACT}}%
  {\ABSTRACT={Not yet given.}}%
  {}%
  {\noindent{\bf Abstract. }\ignorespaces\the\ABSTRACT\vskip0.5true cm}%
\def\strich{\par
\vbox to0pt{\hrule width\hsize\vss}\vskip-1.2\baselineskip
\vskip0pt plus3\baselineskip\relax}%
\tstmiss%
  {\KEYWORDS}%
  {\missmsg{KEYWORDS}}%
  {\KEYWORDS={Not yet given.}}%
  {}%
  {\noindent{\bf Key words: }\ignorespaces\the\KEYWORDS
  \strich}%
\global\sterne=0
}}%Ende von maketitle
\newdimen\@txtwd  \@txtwd=\hsize
\newdimen\@txtht  \@txtht=\vsize
\newdimen\@colht  \@colht=\vsize
\newdimen\@colwd  \@colwd=-1pt
\newdimen\@colsavwd
\newcount\in@t \in@t=0
\def\initlr{\if N\lr \ifdim\@colwd<0pt \global\@colwd=\hsize \fi
   \else\global\let\lr=L\ifdim\@colwd<0pt \global\@colwd=\hsize
      \global\divide\@colwd\tw@ \global\advance\@colwd by -10pt
   \fi\fi\global\advance\in@t by 1}
\def\setuplr#1#2#3{\let\lr=O \ifx#1\lr\global\let\lr=N
      \else\global\let\lr=L\fi
   \@txtht=\vsize \@colht=\vsize \@txtwd=#2 \@colwd=#3
   \if N\lr \else\multiply\@colwd\tw@ \fi
   \ifdim\@colwd>\@txtwd\if N\lr
        \errmessage{The text width is less than the column width}%
      \else
        \errmessage{The text width is less the two times the column width}%
      \fi \global\@colwd=\@txtwd
      \if N\lr\divide\@colwd by 2\fi
   \else \global\@colwd=#3 \fi \initlr \@colsavwd=#3
   \global\@insmx=\@txtht
   \global\hsize=\@colwd}
\def\twocolumns{\@fillpage\eject\global\let\lr=L \@makecolht
   \global\@colwd=\@colsavwd \global\hsize=\@colwd}
\def\onecolumn{\@fillpage\eject\global\let\lr=N \@makecolht
   \global\@colwd=\@txtwd \global\hsize=\@colwd}
\def\newpage{\@fillpage\eject}
\def\@fillpage{\vfill\supereject\if R\lr \null\vfill\eject\fi}

\newbox\@leftcolumn
\newbox\@rightcolumn
\newbox\@outputbox
\newbox\@tempboxa
\newbox\@keepboxa
\newbox\@keepboxb
\newbox\@bothcolumns
\newbox\@savetopins
\newbox\@savetopright
\newcount\verybad \verybad=1010
\def\@makecolumn{\ifnum \in@t<1\initlr\fi
   \ifnum\outputpenalty=\the\verybad1  %%% i.e. 10101 if \verybad=1010
      \if L\lr\else\advance\pageno by1\fi
      \message{Warning: There is a 'widow' line
      at the top of page \the\pageno\if R\lr (left)\fi.
      This is unacceptable.} \if L\lr\else\advance\pageno by-1\fi \fi
   \ifnum\outputpenalty=\the\verybad2
      \message{Warning: There is a 'club' line
      at the bottom of page \the\pageno\if L\lr(left)\fi.
      This is unacceptable.} \fi
   \if L\lr \ifvoid\@savetopins\else\@colht=\@txtht\fi \fi
   \if R\lr \ifvoid\@bothcolumns \ifvoid\@savetopright
       \else\@colht=\@txtht\fi\fi\fi
   \global\setbox\@outputbox
   \vbox to\@colht{\boxmaxdepth\maxdepth
   \if L\lr \ifvoid\@savetopins\else\unvbox\@savetopins\fi \fi
   \if R\lr \ifvoid\@bothcolumns \ifvoid\@savetopright\else
       \unvbox\@savetopright\fi\fi\fi
   \ifvoid\topins\else\ifnum\count\topins>0
         \ifdim\ht\topins>\@colht
            \message{|Error: Too many or too large single column
            box(es) on this page.}\fi
         \unvbox\topins
      \else
         \global\setbox\@savetopins=\vbox{\ifvoid\@savetopins\else
         \unvbox\@savetopins\penalty-500\fi \unvbox\topins} \fi\fi
   \dimen@=\dp\@cclv \unvbox\@cclv % open up \box255
   \ifvoid\bottomins\else\unvbox\bottomins\fi
   \ifvoid\footins\else % footnote info is present
     \vskip\skip\footins
     \footnoterule
     \unvbox\footins\fi
   \ifr@ggedbottom \kern-\dimen@ \vfil \fi}%
}
\def\@outputpage{\@dooutput{\lr}}
\def\@colbox#1{\hbox to\@colwd{\box#1\hss}}
\def\@dooutput#1{\global\topskip=10pt
  \ifdim\ht\@bothcolumns>\@txtht
    \if #1N
       \unvbox\@outputbox
    \else
       \unvbox\@leftcolumn\unvbox\@outputbox
    \fi
    \global\setbox\@tempboxa\vbox{\hsize=\@txtwd\makeheadline
       \vsplit\@bothcolumns to\@txtht
       \makefootline\hsize=\@colwd}%
    \message{|Error: Too many double column boxes on this page.}%
    \shipout\box\@tempboxa\advancepageno
    \unvbox255 \penalty\outputpenalty
  \else
    \global\setbox\@tempboxa\vbox{\hsize=\@txtwd\makeheadline
       \ifvoid\@bothcolumns\else\unvbox\@bothcolumns\fi
       \hsize=\@colwd
       \if #1N
          \hbox to\@txtwd{\@colbox{\@outputbox}\hfil}%
       \else
          \hbox to\@txtwd{\@colbox{\@leftcolumn}\hfil\@colbox{\@outputbox}}%
       \fi
       \hsize=\@txtwd\makefootline\hsize=\@colwd}%
    \shipout\box\@tempboxa\advancepageno
  \fi
  \ifnum \special@pages>0 \s@count=100 \page@command
      \xdef\page@command{}\global\special@pages=0 \fi
  }
\def\balance@right@left{\dimen@=\ht\@leftcolumn
    \advance\dimen@ by\ht\@outputbox
    \advance\dimen@ by\ht\springer@macro
    \dimen2=\z@ \global\the@end=0
    \ifdim\dimen@>70pt\setbox\z@=\vbox{\unvbox\@leftcolumn
          \unvbox\@outputbox}%
       \loop
          \dimen@=\ht\z@
          \advance\dimen@ by0.5\topskip
          \advance\dimen@ by\baselineskip
          \advance\dimen@ by\ht\springer@macro
          \advance\dimen@ by\dimen2
          \divide\dimen@ by2
          \splittopskip=\topskip
          % Now split it to two parts of about the same height
          {\vbadness=10000
             \global\setbox3=\copy\z@
             \global\setbox1=\vsplit3 to \dimen@}%
          \dimen1=\ht3 \advance\dimen1 by\ht\springer@macro
       \ifdim\dimen1>\ht1 \advance\dimen2 by\baselineskip\repeat
       \dimen@=\ht1
       % Restore the column boxes and adjust
       \global\setbox\@leftcolumn
          \hbox to\@colwd{\vbox to\@colht{\vbox to\dimen@{\unvbox1}\vfil}}%
       \global\setbox\@outputbox
          \hbox to\@colwd{\vbox to\@colht{\vbox to\dimen@{\unvbox3
             \vfill\box\springer@macro}\vfil}}%
    \else
       \setbox\@leftcolumn=\vbox{unvbox\@leftcolumn\bigskip
          \box\springer@macro}%
    \fi}
\newinsert\bothins
\newbox\rightins
\skip\bothins=\z@skip
\count\bothins=1000
\dimen\bothins=\@txtht \advance\dimen\bothins by -\bigskipamount
\def\bothtopinsert{\par\begingroup\setbox\z@\vbox\bgroup
    \hsize=\@txtwd\parskip=0pt\par\noindent\bgroup}
\def\endbothinsert{\egroup\egroup
  \if R\lr
    \right@nsert
  \else    % L\lr or N\lr
    \dimen@=\ht\z@ \advance\dimen@ by\dp\z@ \advance\dimen@ by\pagetotal
    \advance\dimen@ by \bigskipamount \advance\dimen@ by \topskip
    \advance\dimen@ by\ht\topins \advance\dimen@ by\dp\topins
    \advance\dimen@ by\ht\bottomins \advance\dimen@ by\dp\bottomins
    \advance\dimen@ by\ht\@savetopins \advance\dimen@ by\dp\@savetopins
    \ifdim\dimen@>\@colht\right@nsert\else\left@nsert\fi
  \fi  \endgroup}
\def\right@nsert{\global\setbox\rightins\vbox{\ifvoid\rightins
    \else\unvbox\rightins\fi\penalty100
    \splittopskip=\topskip
    \splitmaxdepth\maxdimen \floatingpenalty200
    \dimen@\ht\z@ \advance\dimen@\dp\z@
    \box\z@\nobreak\bigskip}}
\def\left@nsert{\insert\bothins{\penalty100
    \splittopskip=\topskip
    \splitmaxdepth\maxdimen \floatingpenalty200
    \box\z@\nobreak\bigskip}
    \@makecolht}
\newdimen\@insht    \@insht=\z@
\newdimen\@insmx    \@insmx=\vsize
\def\@makecolht{\global\@colht=\@txtht \@compinsht
    \global\advance\@colht by -\@insht \global\vsize=\@colht
    \global\dimen\topins=\@colht}
\def\@compinsht{\if R\lr
       \dimen@=\ht\@bothcolumns \advance\dimen@ by\dp\@bothcolumns
       \ifvoid\@bothcolumns \advance\dimen@ by\ht\@savetopright
          \advance\dimen@ by\dp\@savetopright \fi
    \else
       \dimen@=\ht\bothins \advance\dimen@ by\dp\bothins
       \advance\dimen@ by\ht\@savetopins \advance\dimen@ by\dp\@savetopins
    \fi
    \ifdim\dimen@>\@insmx
       \global\@insht=\dimen@
    \else\global\@insht=\dimen@
    \fi}
\newinsert\bottomins
\skip\bottomins=\z@skip
\count\bottomins=1000
\xdef\page@command{}
\newcount\s@count
\newcount\special@pages \special@pages=0
\def\specialpage#1{\global\advance\special@pages by1
    \global\s@count=\special@pages
    \global\advance\s@count by 100
    \global\setbox\s@count
    \vbox to\@txtht{\hsize=\@txtwd\parskip=0pt
    \par\noindent\noexpand#1\vfil}%
    \def\protect{\noexpand\protect\noexpand}%
    \xdef\page@command{\page@command
         \protect\global\advance\s@count by1
         \protect\begingroup
         \protect\setbox\z@\vbox{\protect\makeheadline
                                    \protect\box\s@count
            \protect\makefootline}%
         \protect{\shipout\box\z@}%
         \protect\endgroup\protect\advancepageno}%
    \let\protect=\relax
   }
\def\@startins{\vskip \topskip\hrule height\z@
   \nobreak\vskip -\topskip\vskip3.7pt}
\let\retry=N
\output={\@makecolht \global\topskip=10pt \let\retry=N%
   \ifnum\count\topins>0 \ifdim\ht\topins>\@colht
       \global\count\topins=0 \global\let\retry=Y%
       \unvbox\@cclv \penalty\outputpenalty \fi\fi
   \if N\retry
    \if N\lr     % this is for single column output
       \@makecolumn
       \ifnum\the@end>0
          \setbox\z@=\vbox{\unvcopy\@outputbox}%
          \dimen@=\ht\z@ \advance\dimen@ by\ht\springer@macro
          \ifdim\dimen@<\@colht
             \setbox\@outputbox=\vbox to\@colht{\box\z@
             \unskip\vskip12pt plus0pt minus12pt
             \box\springer@macro\vfil}%
          \else \box\springer@macro \fi
          \global\the@end=0
       \fi
       \ifvoid\bothins\else\global\setbox\@bothcolumns\box\bothins\fi
       \@outputpage
       \ifvoid\rightins\else
       %  Hold \rightins back if there is already a \@savetopins
       \ifvoid\@savetopins\insert\bothins{\unvbox\rightins}\fi
       \fi
    \else
       \if L\lr    % this is the left of two columns
          \@makecolumn
          \global\setbox\@leftcolumn\box\@outputbox \global\let\lr=R%
          \ifnum\pageno=1
             \message{|[left\the\pageno]}%
          \else
             \message{[left\the\pageno]}\fi
          \ifvoid\bothins\else\global\setbox\@bothcolumns\box\bothins\fi
          \global\dimen\bothins=\z@
          \global\count\bothins=0
          \ifnum\pageno=1
             \global\topskip=\fullhead\fi
       \else    % the right column
          \@makecolumn
          \ifnum\the@end>0\ifnum\pageno>1\balance@right@left\fi\fi
          \@outputpage \global\let\lr=L%
          \global\dimen\bothins=\maxdimen
          \global\count\bothins=1000
          \ifvoid\rightins\else
          %  Hold \rightins back if there is already a \@savetopins
             \ifvoid\@savetopins \insert\bothins{\unvbox\rightins}\fi
          \fi
       \fi
    \fi
    \global\let\last@insert=N \put@default
    \ifnum\outputpenalty>-\@MM\else\dosupereject\fi
    \ifvoid\@savetopins\else
      \ifdim\ht\@savetopins>\@txtht
        \global\setbox\@tempboxa=\box\@savetopins
        \global\setbox\@savetopins=\vsplit\@tempboxa to\@txtht
        \global\setbox\@savetopins=\vbox{\unvbox\@savetopins}%
        \global\setbox\@savetopright=\box\@tempboxa \fi
    \fi
    \@makecolht
    \global\count\topins=1000
   \fi
   }
\if N\lr
   \setuplr{O}{\fullhsize}{\hsize}% O = one column
\else
   \setuplr{T}{\fullhsize}{\hsize}% T = two columns
\fi
\def\put@default{\global\let\insert@here=Y
   \global\let\insert@at@the@bottom=N}%
\def\puthere{\global\let\insert@here=Y%
    \global\let\insert@at@the@bottom=N}
\def\putattop{\global\let\insert@here=N%
    \global\let\insert@at@the@bottom=N}
\def\putatbottom{\global\let\insert@here=N%
    \global\let\insert@at@the@bottom=X}
\put@default
\let\last@insert=N
\def\end@skip{\smallskip}
\newdimen\min@top
\newdimen\min@here
\newdimen\min@bot
\min@top=10cm
\min@here=4cm
\min@bot=\topskip
\def\figfuzz{\vskip 0pt plus 6pt minus 3pt}  % more flexible spacing
%--------------------------------------------------------------------
\def\check@here@and@bottom#1{\relax
   \ifvoid\topins\else       \global\let\insert@here=N\fi
   \if B\last@insert         \global\let\insert@here=N\fi
   \if T\last@insert         \global\let\insert@here=N\fi
   \ifdim #1<\min@bot        \global\let\insert@here=N\fi
   \ifdim\pagetotal>\@colht  \global\let\insert@here=N\fi
   \ifdim\pagetotal<\min@here\global\let\insert@here=N\fi
   \if X\insert@at@the@bottom\global\let\insert@at@the@bottom=Y
     \else\if T\last@insert  \global\let\insert@at@the@bottom=N\fi
          \if H\last@insert  \global\let\insert@at@the@bottom=N\fi
          \ifvoid\topins\else\global\let\insert@at@the@bottom=N\fi\fi
   \ifdim #1<\min@bot        \global\let\insert@at@the@bottom=N\fi
   \ifdim\pagetotal>\@colht  \global\let\insert@at@the@bottom=N\fi
   \ifdim\pagetotal<\min@top \global\let\insert@at@the@bottom=N\fi
   \ifvoid\bottomins\else    \global\let\insert@at@the@bottom=Y\fi
   \if Y\insert@at@the@bottom\global\let\insert@here=N\fi }
\def\single@column@insert#1{\relax
   \setbox\@tempboxa=\vbox{#1}%
   \dimen@=\@colht \advance\dimen@ by -\pagetotal
   \advance\dimen@ by-\ht\@tempboxa \advance\dimen0 by-\dp\@tempboxa
   \advance\dimen@ by-\ht\topins \advance\dimen0 by-\dp\topins
   \check@here@and@bottom{\dimen@}%
   \if Y\insert@here
      \par  % The insertion forces a new paragraph in this case.
      \midinsert\figfuzz\relax     %%%%%%%%%\bigskip
      \box\@tempboxa\end@skip\figfuzz\endinsert
      \global\let\last@insert=H
   \else \if Y\insert@at@the@bottom
      \begingroup\insert\bottomins\bgroup\if B\last@insert\end@skip\fi
      \floatingpenalty=20000\figfuzz\bigskip\box\@tempboxa\egroup\endgroup
      \global\let\last@insert=B
   \else
      \topinsert\box\@tempboxa\end@skip\figfuzz\endinsert
      \global\let\last@insert=T
   \fi\fi\put@default\ignorespaces}
\def\begfig#1cm#2\endfig{\single@column@insert{\@startins\rahmen{#1}#2}%
\ignorespaces}
\def\begfigwid#1cm#2\endfig{\relax
   \if N\lr  % Here the only difference to \begfig is the larger \hsize
      {\hsize=\fullhsize \begfig#1cm#2\endfig}%
   \else
      \setbox0=\vbox{\hsize=\fullhsize\bigskip#2\smallskip}%
      \dimen0=\ht0\advance\dimen0 by\dp0
      \advance\dimen0 by#1cm
      \advance\dimen0by7\normalbaselineskip\relax
      \ifdim\dimen0>\@txtht
         \message{|Figure plus legend too high, will try to put it on a
                  separate page. }%
         \begfigpage#1cm#2\endfig
      \else
         \bothtopinsert\line{\vbox{\hsize=\fullhsize
         \@startins\rahmen{#1}#2\smallskip}\hss}\figfuzz\endbothinsert
      \fi
   \fi}
\def\begfigside#1cm#2cm#3\endfig{\relax
   \if N\lr  % Here the only difference to \begfig is the larger \hsize
      {\hsize=\fullhsize \begfig#1cm#3\endfig}%
   \else
      \dimen0=#2true cm\relax
      \ifdim\dimen0<\hsize
         \message{|Your figure fits in a single column; why don't|you use
                  \string\begfig\space instead of \string\begfigside? }%
      \fi
      \dimen0=\fullhsize
      \advance\dimen0 by-#2true cm
      \advance\dimen0 by-1true cc\relax
      \bgroup
         \ifdim\dimen0<8true cc\relax
            \message{|No sufficient room for the legend;
                     using \string\begfigwid. }%
            \begfigwid #1cm#3\endfig
         \else
            \ifdim\dimen0<10true cc\relax
               \message{|Room for legend to narrow;
                        legend will be set raggedright. }%
               \rightskip=0pt plus 2cm\relax
            \fi
            \setbox0=\vbox{\def\figure##1##2{\vbox{\hsize=\dimen0\relax
                           \@startins\noindent\petit{\bf
                           Fig.\ts##1\unskip.\ }\ignorespaces##2\par}}%
                           #3\unskip}%
            \ifdim#1true cm<\ht0\relax
               \message{|Text of legend higher than figure; using
                        \string\begfig. }%
               \begfigwid #1cm#3\endfig
            \else
               \def\figure##1##2{\vbox{\hsize=\dimen0\relax
                                       \@startins\noindent\petit{\bf
                                       Fig.\ts##1\unskip.\
                                       }\ignorespaces##2\par}}%
               \bothtopinsert\line{\vbox{\hsize=#2true cm\relax
               \@startins\rahmen{#1}}\hss#3\unskip}\figfuzz\endbothinsert
            \fi
         \fi
      \egroup
   \fi\ignorespaces}
\def\begfigpage#1cm#2\endfig{\specialpage{\@startins
   \vskip3.7pt\rahmen{#1}#2}\ignorespaces}%
\def\begtab#1cm#2\endtab{\single@column@insert{#2\rahmen{#1}}\ignorespaces}
\let\begtabempty=\begtab
\def\begtabfull#1\endtab{\single@column@insert{#1}\ignorespaces}
\def\begtabemptywid#1cm#2\endtab{\relax
   \if N\lr
      {\hsize=\fullhsize \begtabempty#1cm#2\endtab}%
   \else
      \bothtopinsert\line{\vbox{\hsize=\fullhsize
      #2\rahmen{#1}}\hss}\medskip\endbothinsert
   \fi\ignorespaces}
\def\begtabfullwid#1\endtab{\relax
   \if N\lr
      {\hsize=\fullhsize \begtabfull#1\endtab}%
   \else
      \bothtopinsert\line{\vbox{\hsize=\fullhsize
      \noindent#1}\hss}\medskip\endbothinsert
   \fi\ignorespaces}
\def\begtabpage#1\endtab{\specialpage{#1}\ignorespaces}
\catcode`\@=\active   % This is reset by the \maketitle macro
%
%%%%%%%%%%%%%%%%%%%%%%%%%%%%%%%%%%%%%%%%%%%%%%%%%%%%%%%%%%%%%%%%%%%%5%
\input aa.cmm
  \MAINTITLE={ A forgotten episode of the $\eta ~Car$
light curve in 1860-1865$^*$}
 \FOOTNOTE{  Based on informations collected at the Osservatorio Astronomico
di Bologna historical library (Loiano Observing station)}
\AUTHOR={V.F. Polcaro, R. Viotti }
\INSTITUTE={Istituto di Astrofisica Spaziale, CNR, V. Enrico Fermi 21,
00044 Frascati, Italy}
\KEYWORDS={ history of astronomy; stars: emission-line;
 stars: individual ($\eta ~Car$) }
  \OFFPRINTS={ V.F. Polcaro }
  \DATE={ received February 26; accepted March 10, 1993 }
\THESAURUS={01.08.1; 08.05.2; 08.09.2 $\eta$ Car }

\ABSTRACT={ We have found previously unreported observations of the
galactic LBV $\eta ~Car$ covering the period 1860-1865. Contrary to the
current belief, these data suggest that the star reached the first
magnitude in 1860-1862, with possible large luminosity fluctuations,
followed by a steep fading in 1865.
A revised historical light curve of this
most interesting object is given.}
\maketitle
\titlea{Introduction}
The light history of $\eta ~Car$, one of the few well recognized galactic
Luminous Blue Variables (LBV), can be traced back to the beginning of the
17th century, thanks to the careful discussion of Innes (1903), which was
also summarized by Gratton (1963) and by van Genderen and Th\'e (1984). $\eta
{}~Car$, called $\eta ~Argus$ (or $\eta$ ~of~ the~ $Ship$) in the old
astronomical

literature, was recorded as a fourth magnitude star in the Bayer's Atlas
of 1603 and in the Halley's Catalogue of stars observed at the St. Elena
island (1677), while it was reported as a second magnitude star in the
Lacaille {\it Coelum Stellatum Australis} of 1752, as a fourth magnitude
in 1811-15, and again as a second magnitude
in 1822-27 and in 1828-1832 (Gratton 1963).

In 1827, 1838 and especially 1843, $\eta ~Car$ underwent spectacular
brightenings, during which it attained the first magnitude, being as
bright as $\alpha ~Cru$ (V=0.8) in 1827,
and as $\alpha ~Car$ ($Canopus$, V=--0.7) in 1843.
At that time, $\eta ~Car$ was the brightest star in the sky after $Sirius$
with an apparent magnitude of about --0.8 (Innes 1903). During the 1838 and
1843 maxima, the star brightened by about one magnitude in a few months.
Innes also reports that these maxima were
followed by similar gradual fadings, resembling a kind of large flares or
slow nova-like explosions. After a slow luminosity decrease to the first
magnitude from 1850 to 1855, there was a short-lasting brightening in 1856.
According to Innes (1903), $\eta ~Car$ underwent a spectacular light decrease
from the first to the seventh magnitude.
This deep fading was very slow and lasted about
14 years.
During this phase, the visual luminosity decrease was nearly exponential with
an e-folding time of about 1.9 years. A significant deviation occurred in
1862, when, according the Innes' light curve, the fading phase stopped for
about one year. This behaviour was never considered by previous researchers,
probably because it was considered as one of the many irregularities displayed
by the light curve of this object.

$\eta ~Car$ attained its mimimum luminosity in 1870. Since then its visual
luminosity gradually but irregularly increased until present. This "secular"
trend was marked by a 1 mag outburst in 1889-1893, and by a steeper
brightening in 1940-1950 (O' Connell 1956). Presently, $\eta ~Car$ is a
sixth magnitude star showing small irregular variations.

Because of its big fading, $\eta ~Car$ was in the past included in the
category of novae (indeed the slowest nova even recorded; e.g.
Payne-Gaposchkin 1957; Allen 1973), before being recognized as a completely
different object (an LBV) in more recent years.

\titlea {The Alan Kulczycky 1860-1865 observations}

Despite of the completeness of the $\eta ~Car$ light curve given by Innes
(1903), we have found an original report on the behaviour of this star
concerning the years from 1860 to 1865, which casts some doubts on
the Innes' reconstructed curve for that period.

The data that we have found come from the {\it Connaissance des temps}, a
very authoritative journal published by the French {\it Bureau des
longitudes} since the 18th century, reporting the ephemerides
for professional use for "astronomers and saylors", as well as
until the end of the 19th century, a few papers written
by well known astronomers supposed to be of general interest.
(For instance, the Messier Catalogue of~"nebulous stars" was first published
in this journal).
In the August 1865 issue, reporting the ephemerides for 1867,
this journal published a paper entitled {\it "Observations sur quelques
\'etoiles circompolaires du ciel austral"} ("Observations of some circumpolar
southern stars") by Alan Kulczycky (1865). The aim of the paper was mainly to
correct a number of errors in southern sky catalogues, but
it also included a short note of some variable stars.

Kulczycky starts his note with the description of the
behaviour of $\eta ~Car$ during 1860-65.
We give in the following the text in full and its English translation.

\medskip

{\bf L'\'etoile si remarquable par ses changement d'\'eclat irr\'eguliers,
$\eta$ du Navire, apr\'es avoire brill\'ee comme une \'etoile de $1^{re}$
pendants les ann\'ees 1860, 1861 et 1862, des mani\`ere \`a pouvoire etre
observ\'ee en plein jour, me paraisset cependant diminuer pendant tout cet
intervalle. La dininuition est devenue plus sensible en 1863, au point que
le 20 novembre de cette ann\'ee je n'ai pu l' observer \`a 6$^h$ 46$^m$ du
matin. Pendant l' ann\'ee 1864, en avril, mai, juin, lorsqu'elle se trouvait
au-dessus de l'horizon, elle se d\'etachai bien au milieu de sa n\'ebulosit\'e
et paraissait de 2$^e$ ou de 3$^e$ grandeur. Les nuages me l'ont derob\'ee
ensuite, et il \'etait impossible de suivre exactement ses variations. Mais
en 1865, le 3 avril, j'ai trouve\'e avec \'etonnement que cette \'etoile ne se
distinguait plus dans sa n\'ebuleuse. Le ciel, nuageux, ne m'a permis de la
revoire que dans le 9 avril. Ce jour je l'ai examin\'ee au moyen de la
lunette des passages et d'une lunette \`a pied groissant 56 fois. Je l'ai
compar\'ee aux \'etoiles vicines; {\it p} et {\it q} du Navire, de 4$^e$ et de
5$^e$ grandeur, \'etaint beaucoup plus brillantes, et $\eta$ du Navire me
paraissait exactement \'egale au n$^o$ 3688, estim\'ee de 5$^e$ {\it $1/2$}
grandeur dans le Catalogue, mais que me parait bien de 6$^e$. Depuis cette
\'epoque l'\'etoile me parait changer tr\'es-peu; je crois cependant qu'elle
subit quelque fluctuations peu sensibles dans son \'eclat. Ainsi, le 24 avril,
elle me paraissait plus brillant que le n$^o$ 3688. Je l'ai vue rarement
ensuite \`a cause des nuages; mais elle me parait d\'ecis\'ement diminuer encor
d'\'eclat. Bientot elle ne se trouvera au-dessus de l' horizon que pendant
le jour, et son examen serat n\'ecessairement suspendu}
({\it Connaissance des temps}, August 1865, pp.45-46).

({\it The star $\eta$ of the Ship, very remarkable because of its irregular
change of brigthness, following the brightnening as a 1$^{st}$
magnitude star in 1860,
1861 and 1862, so that it was visible in full day-light, seemed to me to be
fading during all this interval. The fading became more sensible during
1863, so that the 20th November of this year I was unable to observe it at
6$^h$ 45$^m$ a.m.. During 1864, in April, May and June, when it was over
the horizon, it was clearly distiguishable on the middle of its nebulosity
and seemed of 2$^{nd}$ or 3$^{rd}$ magnitude.
In the following, clouds sheathed it to me, and it was impossible to follow
exactly its variations. But, on 1865 April 3rd, I was astonished to find
the star undistinguishable from its nebula. The cloudy sky has not
allowed me to see it again until April 9th. That day, I examined it by means
of the meridian transit instrument
and by means of a portable telescope enlarging 56 times. I compared it with
the nearby stars; p and q of the Ship of 4$^{th}$ and 5$^{th}$ being much more
bright, and $\eta$ of the Ship seemed to me exactly equal to
the star no. 3688, estimated of 5$^{th}1/2$,
but which seemed to me to be of 6$^{th}$.
After this epoch the star seemed to me to change very little even if it
exhibite
d some
minor brightness fluctuations. So, on April 24th, it seemed
brighter than no. 3688.
I saw it rarely thereafter because of the clouds, but
it seemed to me to be still clearly fading its brightness.
Soon it will be below the horizon
but for day-time, and its study will be necessarily interrupted.})

\medskip
It is clear that the aforementioned report contradicts the light curve
reported by Innes (1903).
In fact, according to him the visual magnitude of the star
was always less bright than 1.6 after 1858.5,
and was steady decreasing in brightness from 2.8 in 1860.0 to 5.5
in 1865.3, the period covered by the Kulczycky's observations.
We shall now discuss the possible origin of this contradiction.

We must first notice that the original Innes discussion on the light curve
of $\eta~Car$ is divided into two parts:
the earlier one, up to the beginning of the 19th century, contains
a detailed discussion of the fonts; the second
(covering a 70 years period) is considered based on observations made
following modern standards and thus these observations are not
discussed in details. In fact, Innes states:
\medskip

{\it "The greather part of the observations available are
so good that I have felt it advisable to omit nearly all the numerous meridian
observations of magnitude of the last 70 years. In many cases these do
not pretend to be estimations of magnitude, the magnitudes published in
catalogues being in many cases taken from other sources"}.
\medskip

It is clear that Innes was acquainted with many more
observations than those reported in his paper,
and excluded many of them which on his opinion
were not reliable enough.
In most cases Innes was probably correct,
but we should remind that, at the time of the Innes' work (1903),
there was no knowledge of any stellar
object, excepting novae, which would undergo large luminosity variations
similar to those observed in $\eta ~Car$.
Actually, all the measurements (but two)
reported by Innes for the period 1860-1865 come from only two observers
(Tebbutt and Abbott), and Innes himself mentioned that one of them, Tebbutt,
has later (the end of the century) revised his observations.
It is therefore possible that Tebbutt was convinced that some of his original
measurements, maybe those showing unexpected
peaks of the star brightness, were wrong, and
then these observations were
ruled out in the revision, in order to obtain a smoother curve.
A memory of this possible data revision could be the 1862 stop of the
fading phase reported by the Innes' light curve.
Actually, all the Tebbutt data reported in the Innes paper come from the
same issue (Volume XXXI) of the Monthly Notice of The Royal Astronomical
Society, published much later than the reported observations.
Unfortunately, we were unable to find the original Tebbutt's paper,
which could possibly clarify this point.

\medskip

Let us now examine the report of Kulczycky (1865).
The first consideration is that, contrary to the Innes' one,
this report is based on "first hand"
observations made at the time of its publication.
We have been unable to find any notice about the author, but
useful information can be derived from the paper itself, in which
A. Kulczycky is defined as
"Colonial engineer, and Director of the Port-de-France (New Caledony)
Observatory". It can also be inferred from the paper that he
was a professional astronomer, holding an official
position with the duty of
checking the star maps to be used by the
French Navy. In fact, the instrumentation he used, described as a
"portable meridian transit instrument made by Brunner", cames from the Navy
Storage of charts and maps. Kulczycky's observations should therefore be
quite reliable, and it is very unlikely that his observations would
contain trivial errors, such as the misidentification of
a first magnitude star.
We must also consider that the day-time visibility of
a celestial object is a quite unusual phenomenon
to bring a professional astronomer to mistake another
celestial object (maybe Venus or Sirius) for $\eta~Car$.

Concerning the high luminosity phase of $\eta$ $Car$ during 1860-62,
one could argue that these observations were not made by Kulczycky
himself, as this is not explicitely stated, but that he was
referring to observations made by others. This point could be supported
by the fact that in the footnote of the Kulczycky's
paper it is said that the observations used in that work
were performed (by the author) during 1864-1865 using the
Bruuner portable meridian, and this could be coincident with the
period of permanence of the author in Port-de-France.
On the other hand, in reporting the high luminosity
state of $\eta$ $Car$ during (1860-62), Kulczycky
has added his {\it personal} impression that the star seemed
to be fading during the whole period.
The author also appeared to be astonished by the fact that on November 20,
1863 $he$ was unable to see the star at the sunrise.
We are therefore led to the conclusion that also for the period 1860-63
Kulczycky is reporting his own observations,
which strongly supports the reality of the high luminosity state
of $\eta$ $Car$ during 1860-62.

Fig. 1 shows the light curves of $\eta$ $Car$ based on Innes and
Kulczycky data. For the latter we have assumed a visual magnitude of +1
in 1860, and a slight luminosity decrease in the following 2 years.
The visual magnitudes of 2.5 and 5.5 were adopted
for April-June 1864 and April 1865, respectively.
The slight brightening of April 24th, 1865 is not reported.

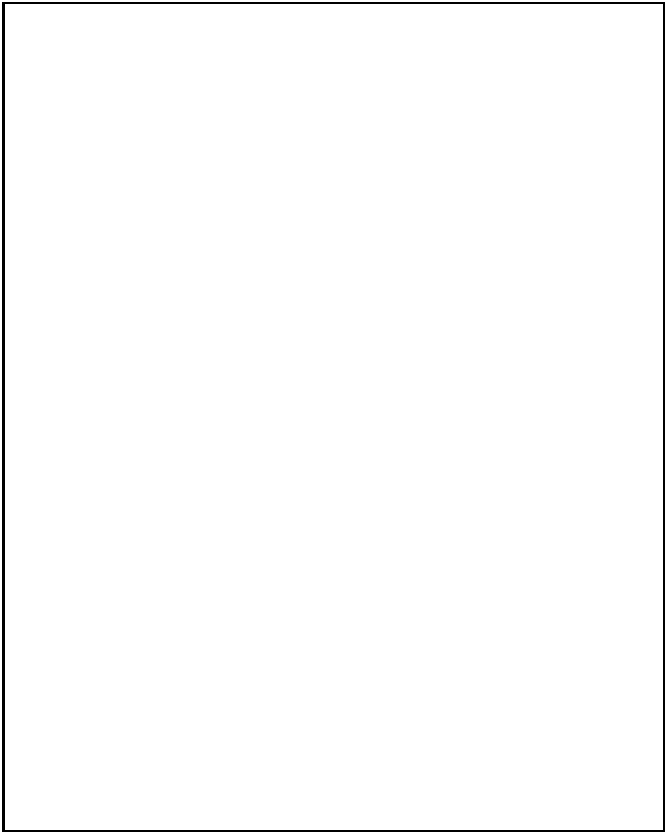
   \begfig 11.0cm \figure{1}{ Dots: the Innes (1903) light curve of
$\eta ~Car$. Triangles: the revised light curve based on the Kulczycky
(1865) paper}
   \endfig

At this point we have to explain why Abbott and Tebbutt have not reported an
extraodinary phenomenon such as the day time visibility of $\eta ~Car$,
as clearly stated by Kulczycky. We should first consider that
Kulczycky does not give the duration of this visibility.
In fact it is possible that the luminosity maximum
of 1860-62 was not a plateau, but a series of short lasting peaks,
which could therefore have been
lost by the other observers for instance because of the bad weather.
In this regard, we remind that, according to Innes (1903)
the 1843 maximum lasted a few days.
The other possibility already discussed above is that some observers
did report these luminosity peaks, but their observations
were later discarded by themselves or by Innes.
A careful analysis of the original observations should be made
in order to clarify this point.

On the other hand, even if we do not consider Kulczycky's data
for 1860-62, there is little doubt about his detailed
description of the star's luminosity during the following period
of 1864-65. According to him the stellar magnitude in 1864 was between
2 and 3, in strong contrast with the value of 4.5 reported by Innes,
while the 1865 magnitudes are in good agreement.
In practice, the slight 1862-63 luminosity hump in the Innes curve
is strongly enhanced, suggesting that either $\eta$ $Car$
underwent a strong outburst (or several short lasting outbursts)
during 1860-63, or the actual stellar luminosity was still the
same as one decade before, but the star was subject to a number of
R CrB-type fadings clearly associated with the formation of
circumstellar dust.

It should be finally remarked that the papers published in the
{\it Connaissance des temps} were widely known to the astronomical
community, certainly including Abbott and Tebbutt,
and that they were frequently followed by discussions and
comments published in the following issues.
This would have certainly be happened if an important announcement
such as the day-time visibility of a star would had been a trivial
mistake. But no comment of the Kulczycky's paper was published
in the following issues of the same journal.

\titlea{Conclusions}

We have found previously unreported observations of $\eta$ $Car$ during
its deep fading phase, which strongly supports the existence of a
more extended high luminosity phase of the star.
The apparent disagreement with the well known light curve of Innes (1903)
could be explained by the presence of large luminosity fluctuations,
which are consistent with the fact that during this period of time the
star was subject to huge mass ejection followed by the formation of
circumstellar dust clouds which were irregularly masking the stellar light.

Clearly, a careful search and analysis of all the archival data on
$\eta ~Car$ is required in order to have a more precise picture
of its historical light curve, in particular before and during
the deep fading phase which is fundamental to unveil
the nature of this extremely interesting object.

\acknow{We would like to aknowledge
the Osservatorio Astronomico di Bologna for the careful
presevation of an enormous amount of ancient astronomical journals,
including a complete collection of the {\it Connaissance des temps},
in the library of the Loiano observing station,
where we have found the data used in this paper. }

\begref {References}
\ref  Allen C.W., 1973 Astrophysical Quantities, third edition, The Athlone
   Press, University of London
\ref  Gratton L., 1963, Star Evolution, in L. Gratton (ed.)
   Academic Press, New York, p.297
\ref Innes R.T.A., 1903, Cp. A. 9, 75B
\ref  Kulczycky A., 1865, {\it Connaissance des temps},
     Bureau des Longitudes, Paris, August, pp. 42-46
\ref  O' Connell D.J.K., 1956, Vistas in Astronomy 2, 1165
\ref  Payne-Gaposchkin C., 1957, Galactic Novae, North Holland
\ref van Genderen A.M., Th\'e P.S., 1984, Space Science Reviews 39, 317
\endref
\bye